\documentclass[submitting]{nst}

\usepackage{subfigure,dcolumn}
\usepackage[T2A,T1]{fontenc}
\usepackage[russian,english]{babel}
\usepackage{hyperref}
\usepackage{ulem}
\usepackage{lineno}

\usepackage{listings}
\begin{document}

\title{Performance of the Gamma-ray Transient Monitor at the IHEP Electron-Beam Facility}


\author{Pei-Yi Feng}
\email[Corresponding author, ]{Pei-Yi Feng, fengpeiyi@ihep.ac.cn.}
\affiliation{State Key Laboratory of Particle Astrophysics, Institute of High Energy Physics, Chinese Academy of Sciences, Beijing 100049, China}
\affiliation{Guangxi Key Laboratory for Relativistic Astrophysics, School of Physical Science and Technology, Guangxi University, Nanning 530004, China}

\author{Zheng-Hua An}
\email[Corresponding author, ]{Zheng-Hua An, anzh@ihep.ac.cn.}
\affiliation{State Key Laboratory of Particle Astrophysics, Institute of High Energy Physics, Chinese Academy of Sciences, Beijing 100049, China}


\author{Yu-Hui Li}
\email[Corresponding author, ]{Yu-Hui Li, liyuhui@ihep.ac.cn.}
\affiliation{Key Laboratory of Particle Acceleration Physics \& Technology, Institute of High Energy Physics, Chinese Academy of Sciences, Beijing 100049, China}

\author{Qi Le}
\affiliation{Key Laboratory of Particle Acceleration Physics \& Technology, Institute of High Energy Physics, Chinese Academy of Sciences, Beijing 100049, China}

\author{Da-Li Zhang}
\affiliation{State Key Laboratory of Particle Astrophysics, Institute of High Energy Physics, Chinese Academy of Sciences, Beijing 100049, China}

\author{Xin-Qiao Li}
\affiliation{State Key Laboratory of Particle Astrophysics, Institute of High Energy Physics, Chinese Academy of Sciences, Beijing 100049, China}

\author{Shao-Lin Xiong}
\affiliation{State Key Laboratory of Particle Astrophysics, Institute of High Energy Physics, Chinese Academy of Sciences, Beijing 100049, China}

\author{Hong-Fei Guan}
\affiliation{Technology and Engineering Center for Space Utilization, Chinese Academy of Sciences, Beijing 100094, China}

\author{Cai-Yun Shao}
\affiliation{Technology and Engineering Center for Space Utilization, Chinese Academy of Sciences, Beijing 100094, China}

\author{Chen-Wei Wang}
\affiliation{State Key Laboratory of Particle Astrophysics, Institute of High Energy Physics, Chinese Academy of Sciences, Beijing 100049, China}

\author{Chao Zheng}
\affiliation{State Key Laboratory of Particle Astrophysics, Institute of High Energy Physics, Chinese Academy of Sciences, Beijing 100049, China}

\author{Jia-Cong Liu}
\affiliation{State Key Laboratory of Particle Astrophysics, Institute of High Energy Physics, Chinese Academy of Sciences, Beijing 100049, China}

\author{Xiang-Yang Wen}
\affiliation{State Key Laboratory of Particle Astrophysics, Institute of High Energy Physics, Chinese Academy of Sciences, Beijing 100049, China}

\author{Sheng Yang}
\affiliation{State Key Laboratory of Particle Astrophysics, Institute of High Energy Physics, Chinese Academy of Sciences, Beijing 100049, China}

\author{Ke Gong}
\affiliation{State Key Laboratory of Particle Astrophysics, Institute of High Energy Physics, Chinese Academy of Sciences, Beijing 100049, China}

\author{Ya-Qing liu}
\affiliation{State Key Laboratory of Particle Astrophysics, Institute of High Energy Physics, Chinese Academy of Sciences, Beijing 100049, China}

\author{Xiao-Jing Liu}
\affiliation{State Key Laboratory of Particle Astrophysics, Institute of High Energy Physics, Chinese Academy of Sciences, Beijing 100049, China}

\author{Min Gao}
\affiliation{State Key Laboratory of Particle Astrophysics, Institute of High Energy Physics, Chinese Academy of Sciences, Beijing 100049, China}

\author{Xiao-Yun Zhao}
\affiliation{State Key Laboratory of Particle Astrophysics, Institute of High Energy Physics, Chinese Academy of Sciences, Beijing 100049, China}

\author{Fan Zhang}
\affiliation{State Key Laboratory of Particle Astrophysics, Institute of High Energy Physics, Chinese Academy of Sciences, Beijing 100049, China}

\author{Jin-Zhou Wang}
\affiliation{State Key Laboratory of Particle Astrophysics, Institute of High Energy Physics, Chinese Academy of Sciences, Beijing 100049, China}

\author{Xi-Lei Sun}
\affiliation{State Key Laboratory of Particle Detection and Electronics, Institute of High Energy Physics, Chinese Academy of Sciences, Beijing 100049, China}

\author{Cong-Zhan Liu}
\affiliation{State Key Laboratory of Particle Astrophysics, Institute of High Energy Physics, Chinese Academy of Sciences, Beijing 100049, China}

\author{Wei-Bin Liu}
\affiliation{Key Laboratory of Particle Acceleration Physics \& Technology, Institute of High Energy Physics, Chinese Academy of Sciences, Beijing 100049, China}

\author{Jian-Li Wang}
\affiliation{Key Laboratory of Particle Acceleration Physics \& Technology, Institute of High Energy Physics, Chinese Academy of Sciences, Beijing 100049, China}

\author{Bing-Lin Deng}
\affiliation{Key Laboratory of Particle Acceleration Physics \& Technology, Institute of High Energy Physics, Chinese Academy of Sciences, Beijing 100049, China}

\author{Yu-Guang Xie}
\affiliation{State Key Laboratory of Particle Detection and Electronics, Institute of High Energy Physics, Chinese Academy of Sciences, Beijing 100049, China}

\author{He Xu}
\affiliation{Institute of High Energy Physics, Chinese Academy of Sciences, Beijing 100049, China}


\author{Hong Lu}
\affiliation{State Key Laboratory of Particle Astrophysics, Institute of High Energy Physics, Chinese Academy of Sciences, Beijing 100049, China}

\begin{abstract}

Gamma-Ray Transient Monitor (GTM) is an all-sky monitor onboard the Distant Retrograde Orbit-A (DRO-A) satellite, with the scientific objective of detecting gamma-ray bursts in the energy range of 20 keV to 1 MeV. GTM is equipped with five Gamma-Ray Transient Probes (GTPs), utilizing NaI(Tl) scintillators coupled with silicon photomultiplier (SiPM) arrays for signal readout. To test the performance of the GTP in detecting electrons, we used the IHEP Electron-Beam Facility (a continuous-energy-tunable, low-current, quasi-single-electron accelerator) for ground-based electron tests of the GTP. This paper provides a detailed description of the operating principles of the electron accelerator and presents the process and results of the GTP electron-beam tests. The test results show that the GTP has a dead time of less than 4 $\mu$s for normal signals and approximately 70 $\mu$s for overflow signals, consistent with the design specifications. The time-recording capability of the GTP was tested and found to be normal, with accurate recording of overflow events. The GTP's response to electrons in the 0.4-1.4 MeV range is also normal. Additionally, we used Geant4 to simulate the GTP's energy response and performed a comparative analysis of the simulation and experimental results. The performance tests and ground-based electron calibration validated the design of the GTP and enhanced the GTP's mass model, laying the foundation for payload development, in-orbit observation strategies, and scientific data analysis.

\end{abstract}




\maketitle
\nolinenumbers

\section{Introduction}


The first observation of the electromagnetic counterpart event GRB170817A associated with gravitational waves (GWs) marked a significant breakthrough, heralding the era of "multi-messenger, multi-wavelength" astronomy \cite{abbott2017gw170817, abbott2017multi, abbott2017gravitational, li2018insight}. Since May 2023, ground-based gravitational wave detectors (LIGO, Virgo, and KAGRA) have commenced a new phase of scientific observation (O4), providing unprecedented opportunities to discover high-energy electromagnetic counterparts related to gravitational waves \cite{arnaud2023ligo, shah2024predictions}. 


Several dedicated instruments are used to detect gamma-ray transients, such as Konus-Wind \cite{aptekar1995konus}, CGRO/BATSE \cite{gehrels1993compton}, Swift \cite{gehrels2004swift}, Fermi/GBM \cite{meegan2009fermi}, MAXI \cite{mihara2014maxi}, GECAM \cite{an2022design, li2021technology, feng2024detector}, EP \cite{yuan2022einstein}, SVOM \cite{atteia2022svom}, and CATCH \cite{li2023catch}. However, most instruments operate in low-Earth orbit (LEO), where they pass through the South Atlantic Anomaly (SAA) or high-latitude radiation zones, resulting in a complex space environment background. The Earth's magnetic field protection is relatively weak over the SAA region, allowing more radiation from outer space to penetrate and reach lower altitudes of the atmosphere, which can easily disrupt communications for satellites, aircraft, and spacecraft passing through the SAA. Additionally, LEO satellites face inevitable Earth occultation. These are significant disadvantages of LEO instruments. To better observe gamma-ray transients, we strongly hope to deploy detection instrument in deep space.



\begin{figure*}[!htb]
\includegraphics[width=\hsize]
{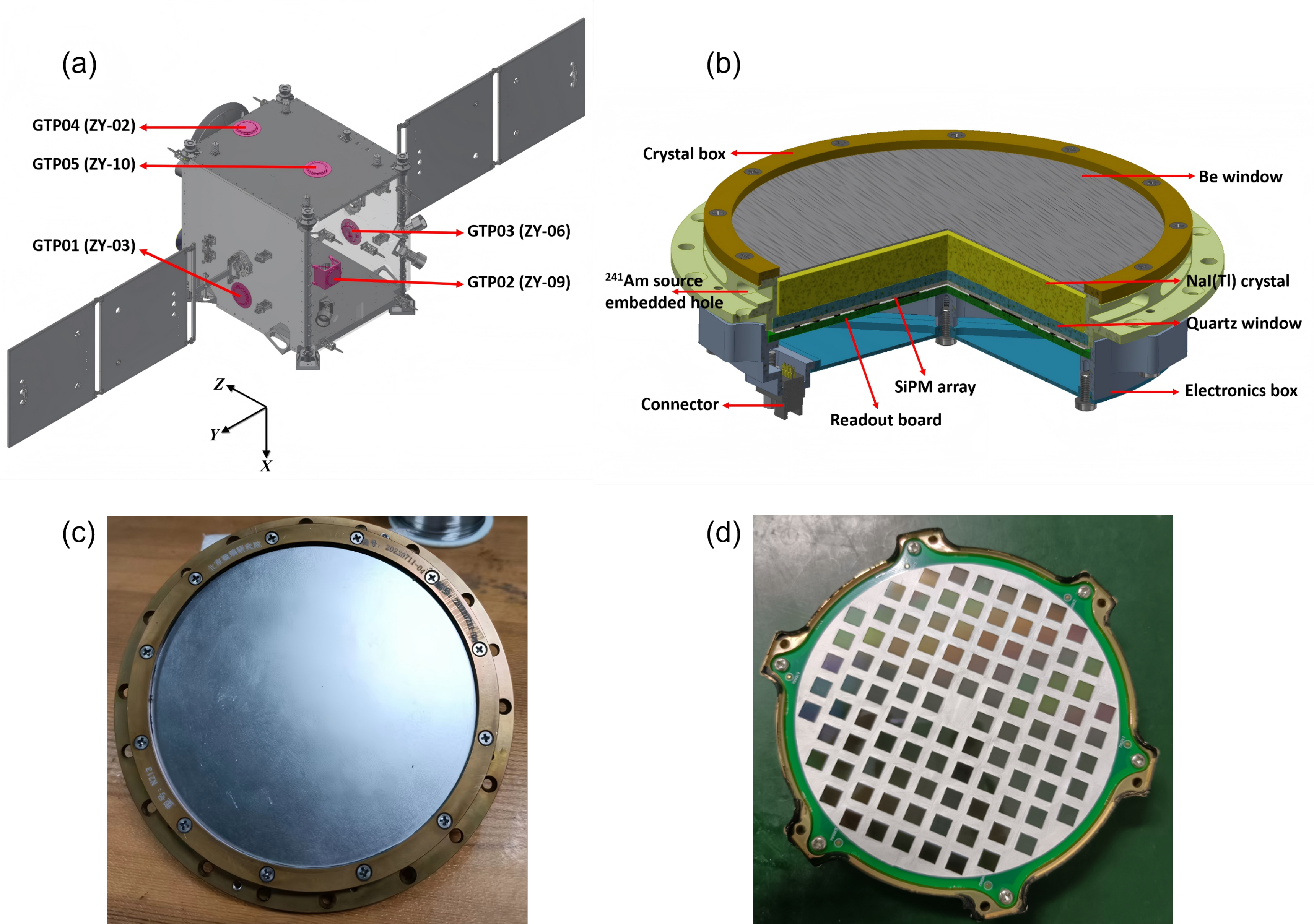}
\caption{(a) Overview of DRO-A satellite. GTM consists of five Gamma-ray Transient Probes (GTPs) positioned on the four sides of the spacecraft. Four standard GTPs are individually mounted on the ±Y side (one GTP for each side) and the –X side (two GTPs), while one dedicated GTP for the –Z side. The standard GTP comprises detector components and radiation cooling plates, while the dedicated GTP is composed of detector components and brackets (with the brackets also doubling as radiation cooling plates). The detector label with GTP is for science usage, whereas that with ZY is for the crystal label of detector \cite{feng2024detector}. (b) Structural diagram of GTP module onboard GTM \cite{feng2024detector}. (c) NaI(Tl) crystal encapsulation. (d) A 100-chip SiPM array and a preamplifier circuit board.}
\label{fig:1}
\end{figure*}

The primary scientific objective of the Gravitational wave burst high-energy Electromagnetic Counterpart All-sky Monitor (GECAM) is to detect and locate gamma-ray transients, particularly those associated with GWs and FRBs \cite{zhang2022dedicated, zhang2019energy, zhang2022gain, zhao2023gecam}. GECAM-A, GECAM-B, and GECAM-C (also called High Energy Burst Searcher, HEBS) have been successfully launched into LEO, achieving a series of significant scientific discoveries \cite{zhang2023performance, sun2023magnetar, FPY}. The most significant achievement is that GECAM-C detected the brightest gamma-ray burst to date, GRB20221009A, without any data saturation or other instrument-related artifacts \cite{an2023insight, zheng2024observation, zhang2024relation}. 

To further enhance the joint monitoring and localization capabilities for gamma-ray transients, we proposed installing a Gamma-ray Transient Monitor (GTM) on the Distant Retrograde Orbit-A (DRO-A) spacecraft (Fig.~\ref{fig:1} (a)), which was launched in March 2024 \cite{feng2024detector, wang2024simulation}. GTM, also known as GECAM-D, inherits the hardware, software, and scientific operations of GECAM mission. Similar to the previous GECAM payloads \cite{SSPMA-2019-0417, SSPMA-2020-0457}, GTM's primary scientific objectives are to monitor high-energy electromagnetic counterparts associated with GWs, GRBs, SGRs, high-energy emission associated with FRBs, and other high-energy transient events. The GTM operates in a deep space orbit, offering advantages such as an unobstructed field of view free from nearby celestial body interference and a more stable space environment, making it highly valuable for monitoring high-energy transient events.


GTM is equipped with five Gamma-ray Transient Probes (GTPs), designed to detect gamma-ray transients in the energy range of 20 keV to 1 MeV. The installation positions of these five GTPs are shown in Fig.~\ref{fig:1} (a). Comprehensive ground tests are essential to ensure that GTPs accurately measure the spectral information of gamma-ray transient sources. These tests aim to characterize the detector's energy response, which is necessary for the reconstruction of in-flight particle flux. The ground test project for X-rays and gamma rays includes aspects such as coincidence time, energy response, detection efficiency, spatial response, bias-voltage response, and temperature dependence, all of which have been completed \cite{feng2024detector}. Electron test is also an important component of ground tests. Although the GTM's orbit does not pass through the SAA, it encounters a different space-radiation environment during deep-space operation, particularly when passing through the magnetotail \cite{nishida2000earth}, where electron and proton activities significantly increase, potentially contributing to the background spectrum \cite{wang2024simulation}. To test the performance of the GTP in this environment, ground electron test using high-energy electron beams is necessary, focusing primarily on two aspects: the dead-time testing in the 0.4-20 MeV range and the energy spectrum in the 0.4-1.4 MeV range.


There are several operational electron-beam facilities internationally. The DA$\Phi$NE Beam Test Facility (BTF) provides an $e^-/e^+$ beam with an energy range of 30-750 MeV, a pulse intensity from a single particle to 10$^{10}$ particles, and a maximum pulse repetition rate of 50 Hz \cite{valente2006diagnostics}. The Medical and Industrial Radiation Facility (MIRF) at the U.S. National Institute of Standards and Technology (NIST) features a linear electron accelerator that delivers electron beams in the energy range of 7-32 MeV, with intensities from a few nA to several tens of $\mu$A \cite{bateman2003nist}. The Joint Institute for Nuclear Research (JINR) in Russia operates the LINAC-200 linear accelerator, whose first section can provide electrons in the energy range of 10-25 MeV \cite{krmar2019beam}. These electron accelerators have been used to calibrate numerous space particle detectors \cite{raggi2017performance, ambrosi2020beam}.

To study the performance of the GTP under electron beam conditions, we tested the GTP using a high-energy electron accelerator. This electron accelerator is located at the Institute of High Energy Physics, Chinese Academy of Sciences (IHEP, CAS), and is characterized by its low current and wide energy range. We refer to it as the IHEP Electron-Beam Facility. This paper first provides a detailed description of the GTP design and the operating principles of the electron accelerator. It then presents performance tests of the GTP using the electron accelerator and simulations of the GTP's energy response to electrons using Geant4. Finally, it compares the results from the simulations and experiments. Ground electron tests not only validated the GTP's mass model and Monte-Carlo-simulated energy response but also laid the foundation for establishing a calibration database.

\section{Instrument Design and Experimental Setup}
In this work, we used the IHEP Electron-Beam Facility to conduct performance tests of the Gamma-ray Transient Probe (GTP). This section introduces the design of the GTP and the operating principles of the IHEP Electron-Beam Facility.

\subsection{Design of Gamma-ray Transient Probe}\label{chap:Design of Gamma-ray Transient Prob}
 
A single GTP (Fig.~\ref{fig:1} (b)) installed on the GTM payload consists mainly of a NaI(Tl) crystal coupled with a 100-chip SiPM array. The NaI(Tl) crystal, with a diameter of 115 mm and a thickness of 10 mm, serves as the sensitive detection material for the GTP. For GRBs of the same brightness in identical space environments, the effective area of the GTP directly determines its sensitivity. When the number of GTPs is limited by factors such as weight, power consumption, and electronics complexity, a sufficiently large area for a single GTP can ensure the detection sensitivity for GRBs. Since NaI(Tl) crystal is hygroscopic, it must be well encapsulated. Figure~\ref{fig:1} (b) shows a schematic of the encapsulated GTP structure. Figures~\ref{fig:1} (c) and (d) present the packaged NaI(Tl) crystal and the SiPM readout array with the preamplifier circuit board, respectively. The SiPM is a highly sensitive photodetector that offers several advantages, including compactness, light weight, low power consumption, and the elimination of high voltage requirements. It has been successfully implemented in the gamma-ray detectors (GRD) on GECAM-A/B and GECAM-C \cite{zhang2023performance, zhang2022quality, zhang2019energy, zhang2022dedicated}.

One of the features of the GTP is the inheritance and development of GRD technology \cite{feng2024sipm, feng2024detector}, using a 4.5-inch circular SiPM array instead of traditional photomultiplier tube (PMT). The SiPM array, consisting of 100 SiPM chips uniformly arranged in a circular pattern, has reflective membranes filling the gaps to improve light collection uniformity. The SiPM array is divided into two groups, each with parallel output and an independent readout system. Offline data processing can time-coincide data from these two channels to eliminate SiPM dark noise, which is a highlight of the GTP design. Previous research concluded that a coincidence time window of 0.5-1 $\mu$s is appropriate \cite{feng2024detector}. Therefore, for the data processing involved in this paper, a coincidence time window of 0.5 $\mu$s is adopted. 

Gamma rays enter through a 400 $\mu$m-thick Be window and a 600 $\mu$m-thick Teflon reflective material, interact with the NaI(Tl) crystal, and the resulting scintillation light passes through a 3 mm-thick quartz glass window to be received by the SiPMs. The SiPMs convert the light signal into an electrical signal, which is amplified by a preamplifier and connected via electrical connectors to the data acquisition system for recording. The GTP provides both energy and time information of the gamma rays for physical analysis. Additionally, a hole on the side of the GTP is designed to embed a $^{241}$Am radioactive source with an activity range of 500 to 800 Bq, whose decay generates a 59.5 keV gamma-ray full-energy peak used for in-orbit calibration of the GTM instrument. The design of the GTP is discussed in greater detail in previously published article \cite{feng2024detector}.

\subsection{IHEP Electron-Beam Facility}
\label{chap:High-energy Electron Accelerator Calibration Facility}

To meet the calibration requirements of high-energy particle detectors, particularly space particle detectors, the Institute of High Energy Physics (IHEP) developed a continuous-energy-tunable, low-current, quasi-single-electron high-energy accelerator, known as the IHEP Electron-Beam Facility. The electron beam is generated by accelerating low-energy electrons emitted from a low-current electron gun. It can provide pulsed beams with intensities ranging from a single electron to several tens of electrons, with energies ranging from 100 keV to 50 MeV. This system can be used for the testing and calibration of various detectors, such as silicon detectors, scintillator counters, and crystal detectors. In this work, we used this electron-beam facility for detailed performance testing of the GTP.


\begin{figure*}[!htb]
\includegraphics[width=0.8\hsize]
{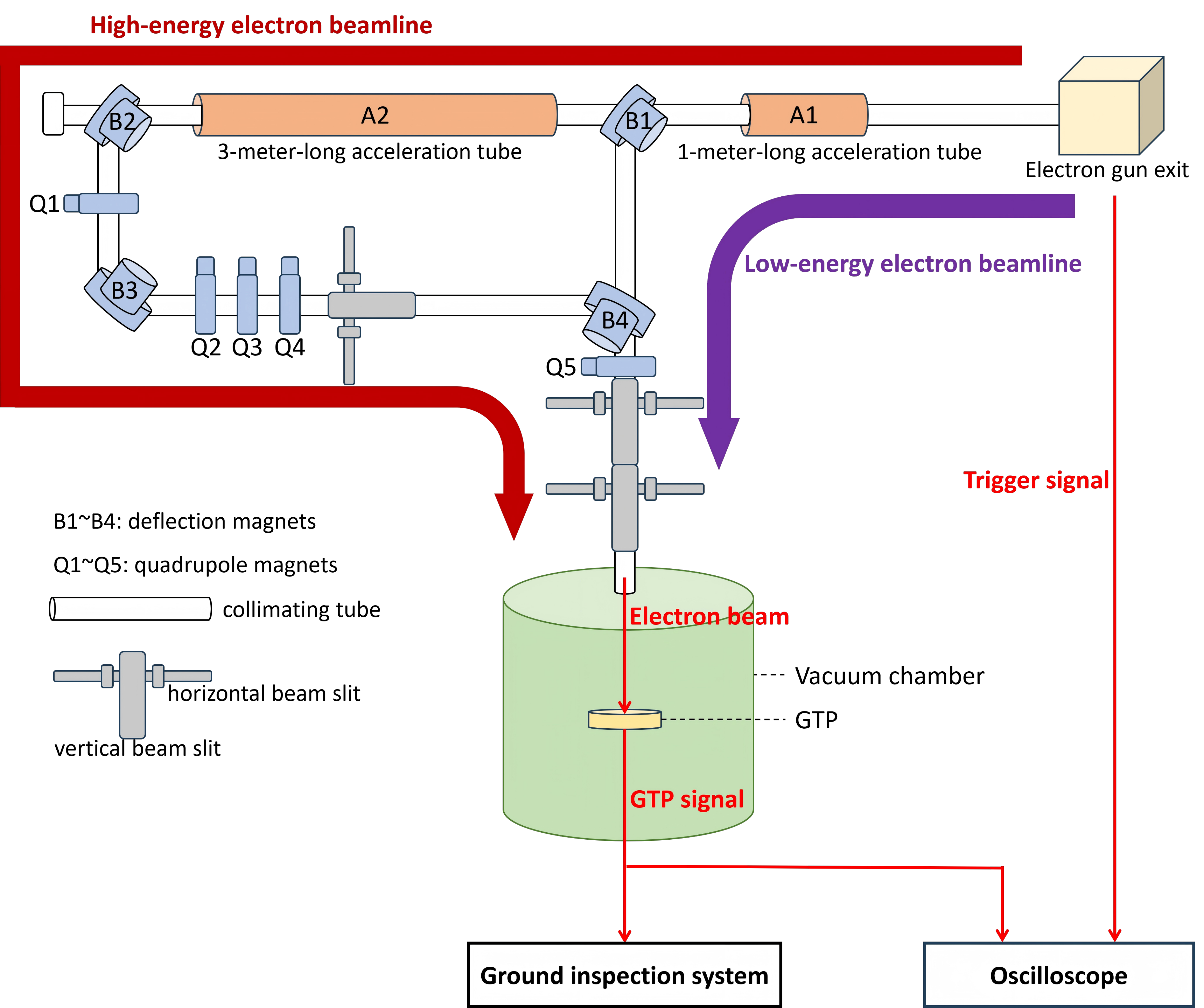}
\caption{Experimental layout of the continuous-energy-tunable, low-current, quasi-single-electron accelerator, located at the Institute of High Energy Physics, Chinese Academy of Sciences (IHEP Electron-Beam Facility).}
\label{fig:2}
\end{figure*}

\begin{figure*}[!htb]
\includegraphics[width=\hsize]
{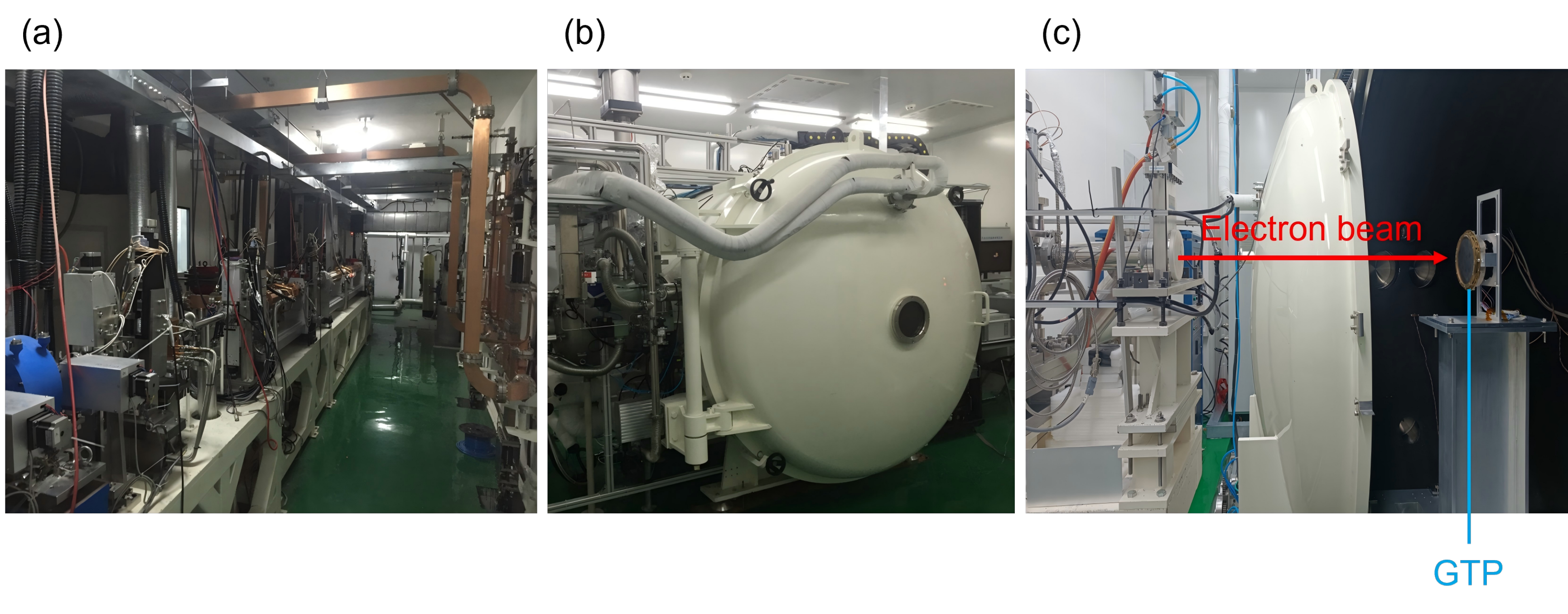}
\caption{(a) Electron accelerator tunnel. (b) Large vacuum chamber. (c) Electron testing site. The GTP is placed inside the vacuum chamber, and a laser level is used to align the positions of the GTP and the beam, with a positional deviation of approximately 2 mm between the center of the electron beam and the GTP.}
\label{fig:3}
\end{figure*}

Figure~\ref{fig:2} shows the experimental layout of the IHEP Electron-Beam Facility. To achieve beam energies adjustable from 100 keV to 50 MeV, the facility's acceleration tubes are divided into two sections: a 1-meter-long tube (A1) and a 3-meter-long tube (A2). Following the 1-meter-long acceleration tube is a deflection magnet (B1) that directs low-energy beams to the side terminal of the accelerator. The low-energy range can cover beams with energies below 7 MeV. High-energy beams exceeding 7 MeV need to pass through the 3-meter-long acceleration tube after the 1-meter tube for further acceleration, then consecutively through deflection magnets B2 to B4, finally reaching the side terminal of the accelerator. The accelerator terminal is connected to a vacuum chamber, within which the GTP is positioned. Signals are transmitted through cables to an external data acquisition system (Fig~\ref{fig:2}). Figures~\ref{fig:3} (a), (b), and (c) show the physical images of the electron accelerator tunnel, the large vacuum chamber, and the GTP testing site, respectively.

\begin{figure*}[!htb]
\includegraphics[width=\hsize]
{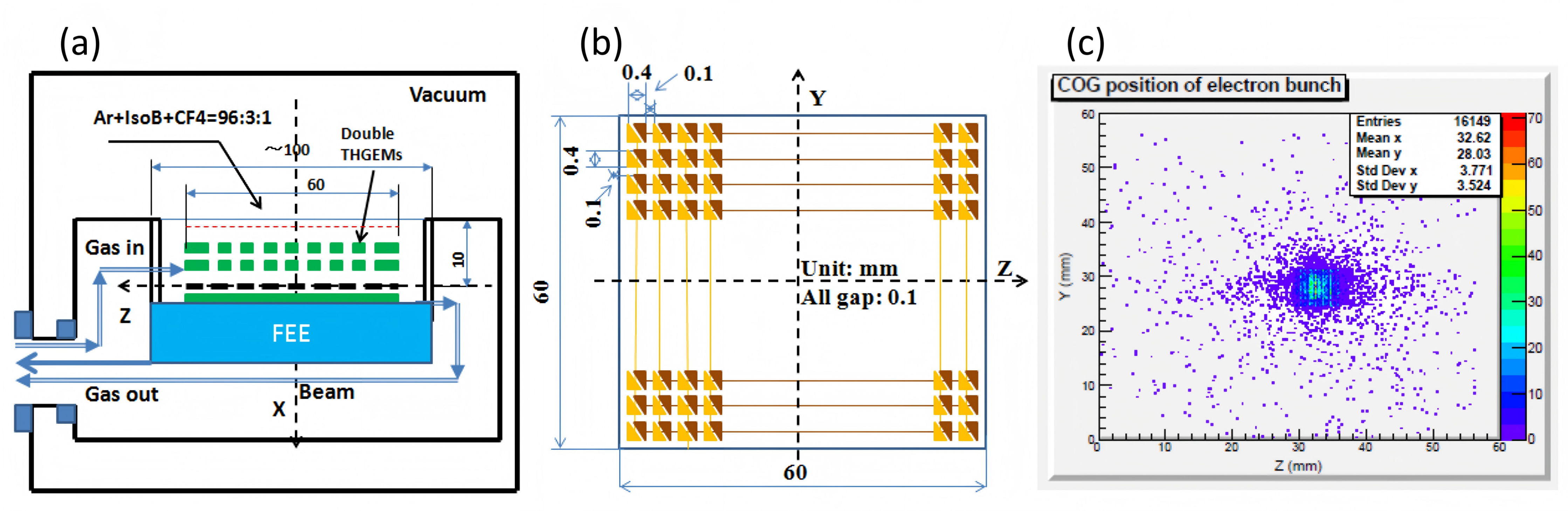}
\caption{(a) Composition and schematic diagram of the particle distribution detector (PDD) system. (b) Readout plane of the particle distribution detector. (c) Transverse distribution of the centroid position of the 40 MeV electron beam, based on the two-dimensional spatial monitoring results from the particle distribution detector.}
\label{fig:4}
\end{figure*}

\begin{figure*}[!htb]
\includegraphics[width=0.7\hsize]
{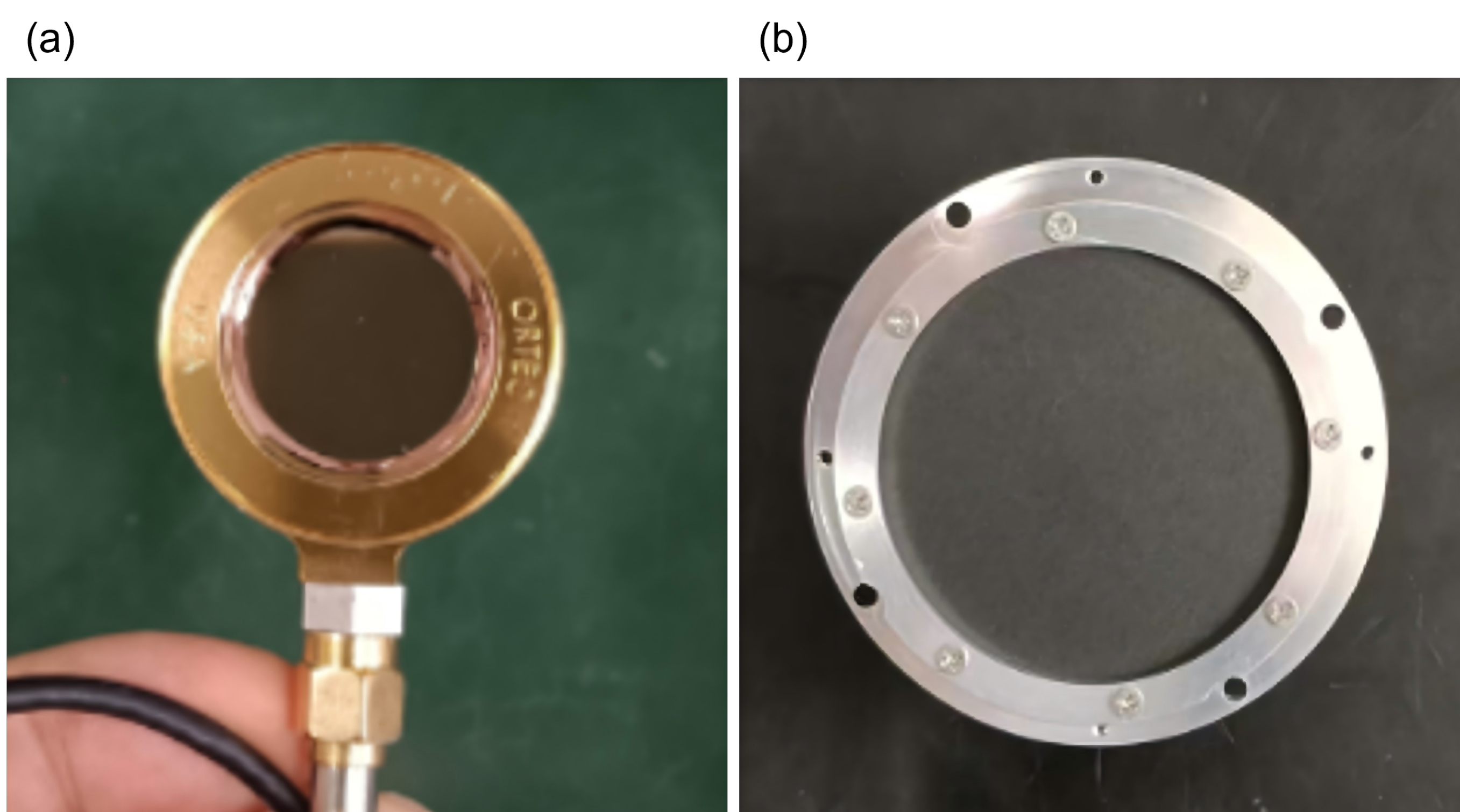}
\caption{Silicon detector (a) and plastic scintillator detector (b) are used to monitor beam energy and electron count.}
\label{fig:5}
\end{figure*}

\begin{figure}[!htb]
\includegraphics[width=\hsize]
{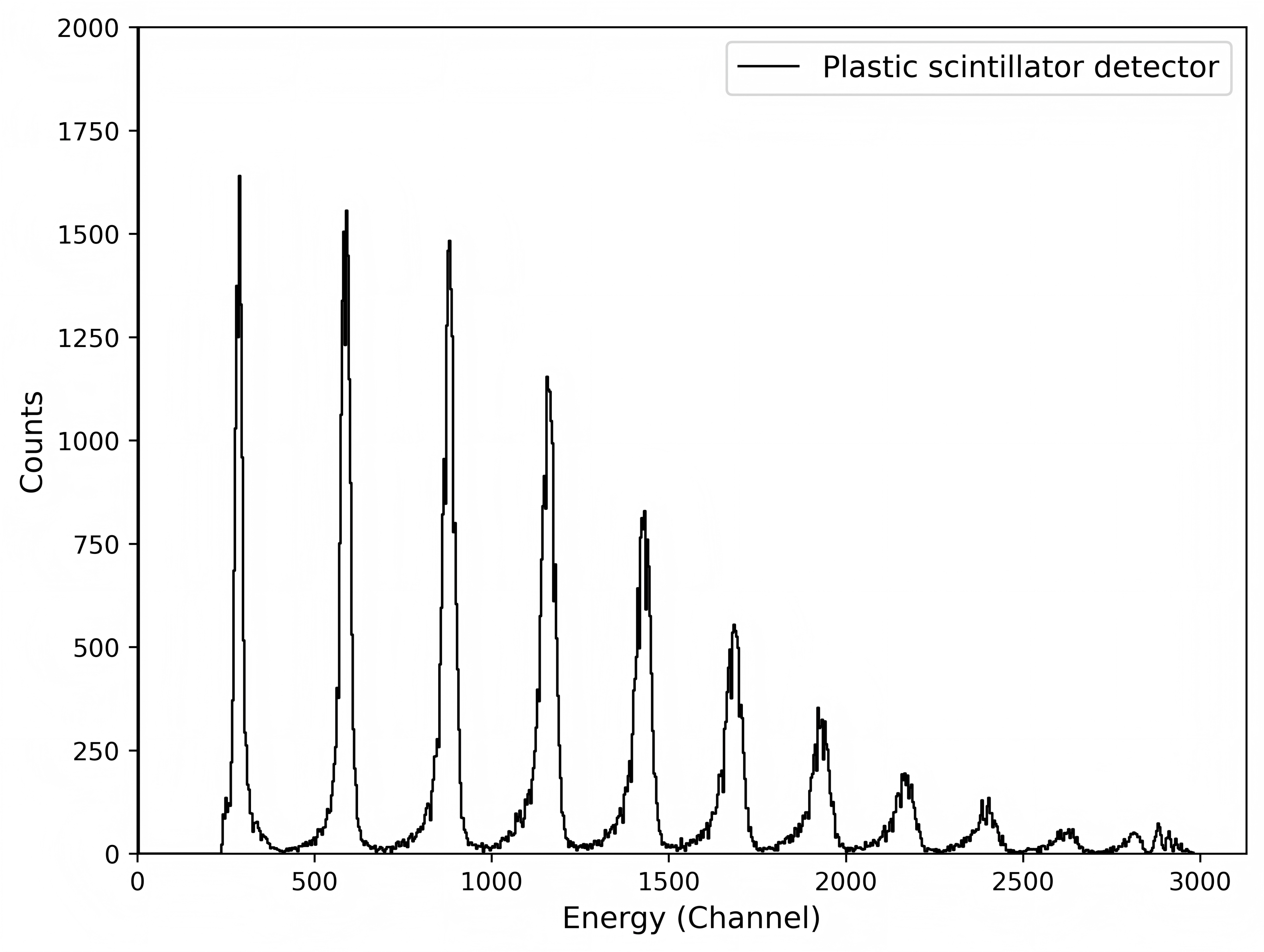}
\caption{Multi-electron spectrum of the 2.5 MeV electron beam measured by the plastic scintillator detector.}
\label{fig:6}
\end{figure}

To obtain quasi-single-electron beams, beam collimators are placed along the accelerator and at the terminal. The attenuation factor of the collimator slits, measured under normal electron beam conditions, is approximately 1/350. After the quadrupole magnet Q4 in the high-energy section, one set of horizontal and vertical beam slits is installed. After the final quadrupole magnet Q5, two sets of horizontal and vertical beam slits are installed to scrape the beam according to its angle and transverse position.
The control of beam energy and energy spread at the accelerator terminal is achieved using the final 90-degree deflection magnet and the beam collimators at the end of the accelerator. Ideally, the relationship between the beam's bending radius in the magnetic field, the magnetic field strength, and the beam energy is given by Equation~\ref{eq:Calibration Facility}.

\begin{equation}\label{eq:Calibration Facility}
\frac{1}{\rho}[m^{-1}] = \frac{eB_0}{p} = 0.2998\frac{B_0[T]}{p[GeV/c]}
\end{equation}

Here, $\rho$ represents the electron's bending radius, $B_0$ is the magnetic field strength, and $p$ is the electron's momentum. To achieve accurate control of beam energy, the magnetic field distribution and excitation curve of the deflection magnet were precisely measured. This data was then used to calculate the magnet's excitation current for different beam energies, followed by particle tracking calculations for further refinement. These steps also determine the relationship between the collimator slit and the beam energy spread.



The beam distribution at the accelerator's terminal is monitored using a particle distribution detector (PDD), as shown in Fig~\ref{fig:4}(a). This detector employs thick gas electron multiplier (THGEM) technology to measure the two-dimensional spatial position of quasi-single electron beams. After the electron beam passes through a thin film window and enters the drift electric field region, ionization occurs. The ionized electrons are then amplified through multiple stages and collected by the induction readout plane. The readout plane is divided into 100 signal readout strips in both the Z and Y directions (Fig~\ref{fig:4}(b)), totaling 200 channels. Each Z and Y readout strip has a width of 0.46 mm and a spacing of 0.1 mm. The detector's data acquisition system measures the charge amount in each channel, determining the beam's incident position using the centroid algorithm. This system can perform online profile monitoring of ultra-low intensity beams. The position where the quasi-single electron beam strikes the PDD detector is uncertain. Its spatial distribution range is related to the beam's energy, count rate, average number of electrons, and the slit width. Figure~\ref{fig:4}(c) shows the centroid distribution of quasi-single electrons striking the PDD at an energy of 40 MeV, with an average of approximately 1.5 electrons. It can be seen that the width of the beam centroid variation is about 20 mm, and the root mean square (RMS) of the centroid variation for every 1000 triggers is about 4 mm.

Silicon detectors and plastic scintillator detectors are used at the terminal of the electron accelerator beam to monitor beam energy and electron count. The energy measurement range of the silicon detector is 100 keV to 2 MeV, while the range of the plastic scintillator detector is 500 keV to 5 MeV. Figure~\ref{fig:5} shows the silicon detector and plastic scintillator detector used for beam energy monitoring. Figure~\ref{fig:6} presents the energy spectrum of a 2.5 MeV electron beam measured with the plastic scintillator detector, clearly showing the characteristic multi-electron spectral shape.

\section{Electron Energy Response Simulation}\label{chap:Electron Energy Response Simulation}


\begin{figure*}[!htb]
\includegraphics[width=\hsize]
{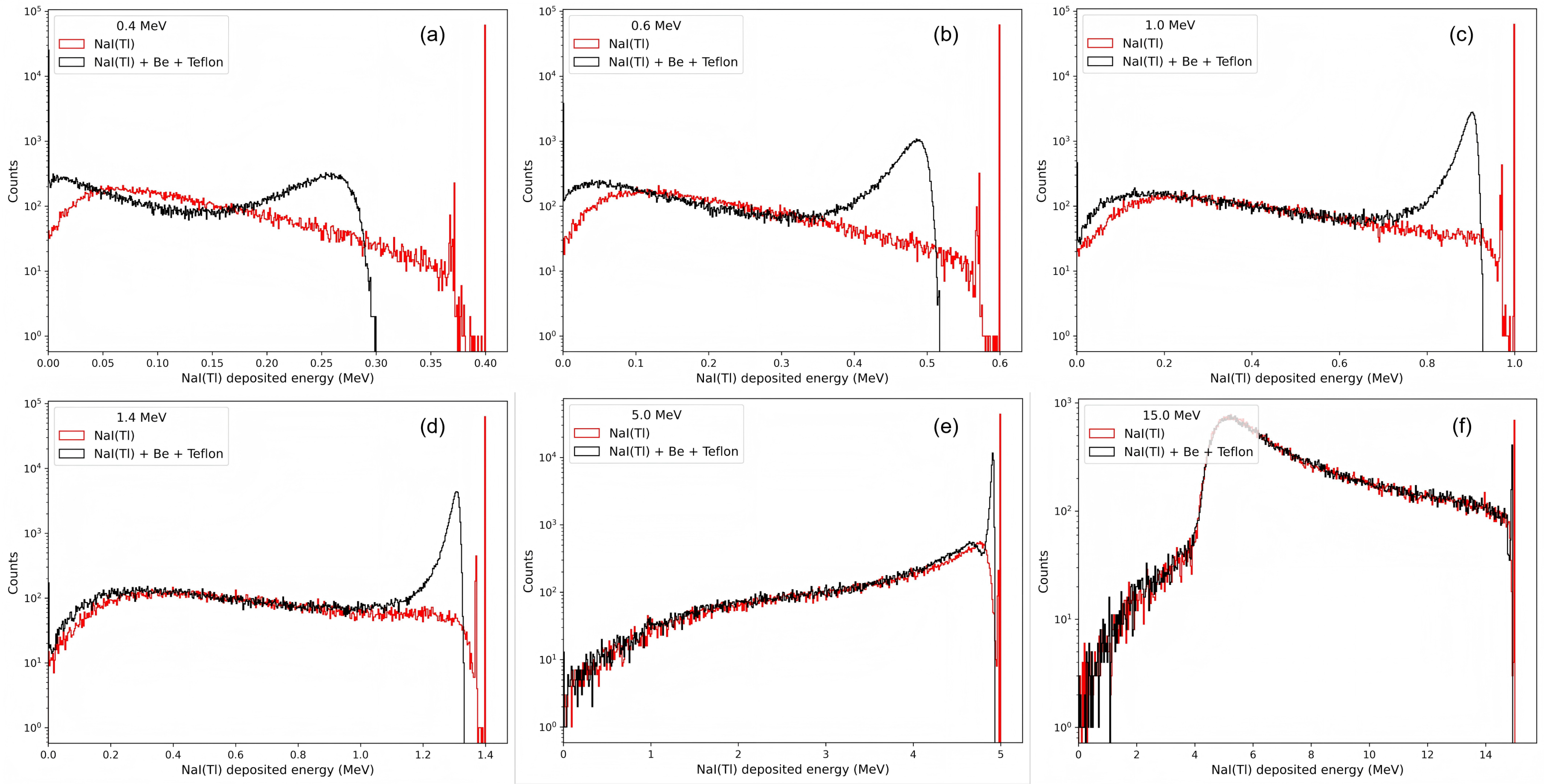}
\caption{Deposited energy spectra detected by the GTP in two scenarios: with and without the Be window and Teflon material. The incident electron energies are 0.4 MeV (a), 0.6 MeV (b), 1.0 MeV (c), 1.4 MeV (d), 5.0 MeV (e), and 15.0 MeV (f).}
\label{fig:15}
\end{figure*}


\begin{figure*}[!htb]
\includegraphics[width=\hsize]
{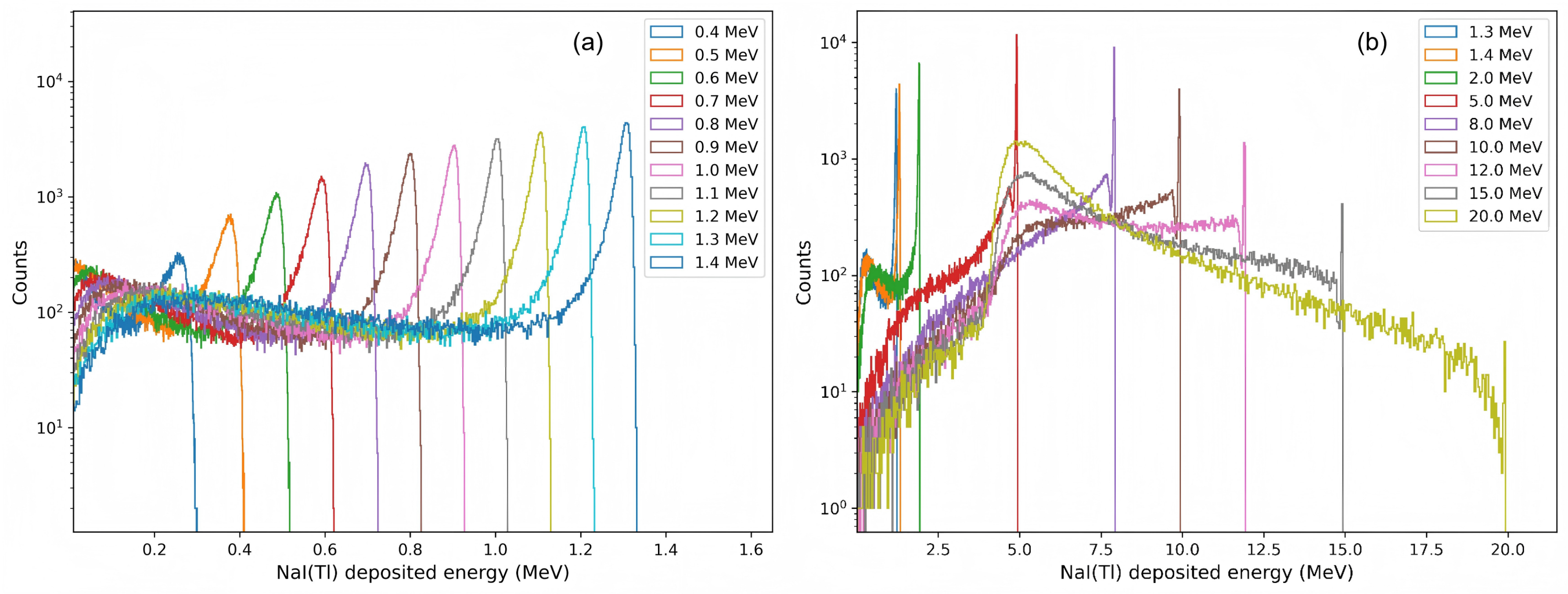}
\caption{Deposited energy spectra of GTP (with the Be window and Teflon) for electrons in the low incident energy range of 0.4-1.4 MeV (a) and high incident energy range of 1.3-20 MeV (b).}
\label{fig:11}
\end{figure*}

\begin{figure*}[!htb]
\includegraphics[width=\hsize]
{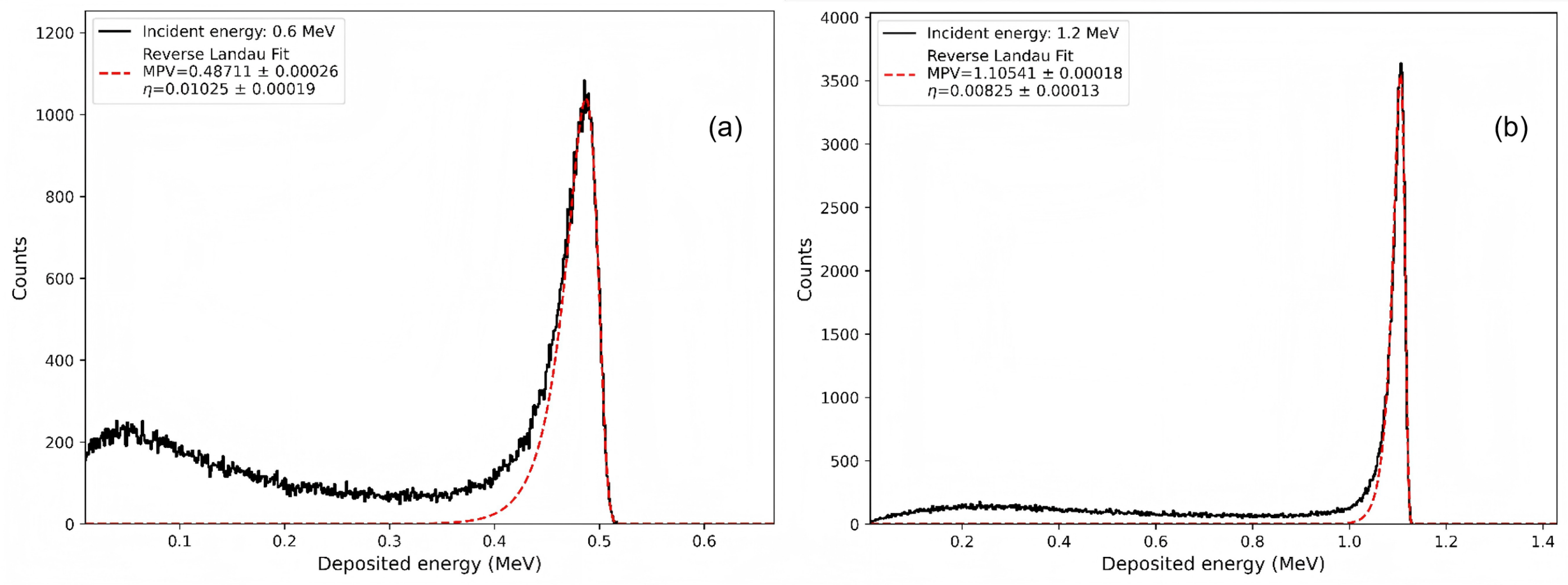}
\caption{Reverse-Landau fit results of the deposited energy spectra for GTP with incident electron energies of 0.6 MeV (a) and 1.2 MeV (b) are presented, where the MPV represents the most probable energy deposited in the GTP.}
\label{fig:17}
\end{figure*}

\begin{figure*}[!htb]
\includegraphics[width=\hsize]
{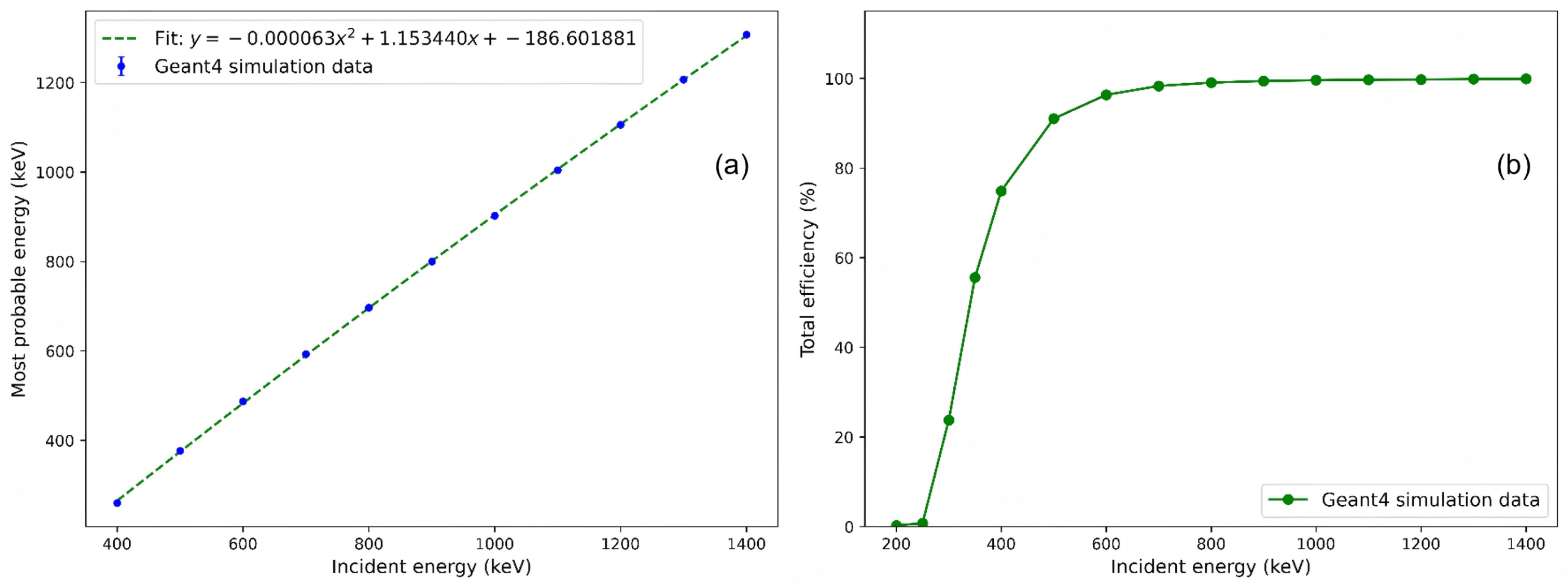}
\caption{(a) Relationship between the incident electron energy and the most probable energy. (b) Geant4 simulation results of the GTP's detection efficiency.}
\label{fig:18}
\end{figure*}

\begin{figure*}[!htb]
\includegraphics[width=\hsize]
{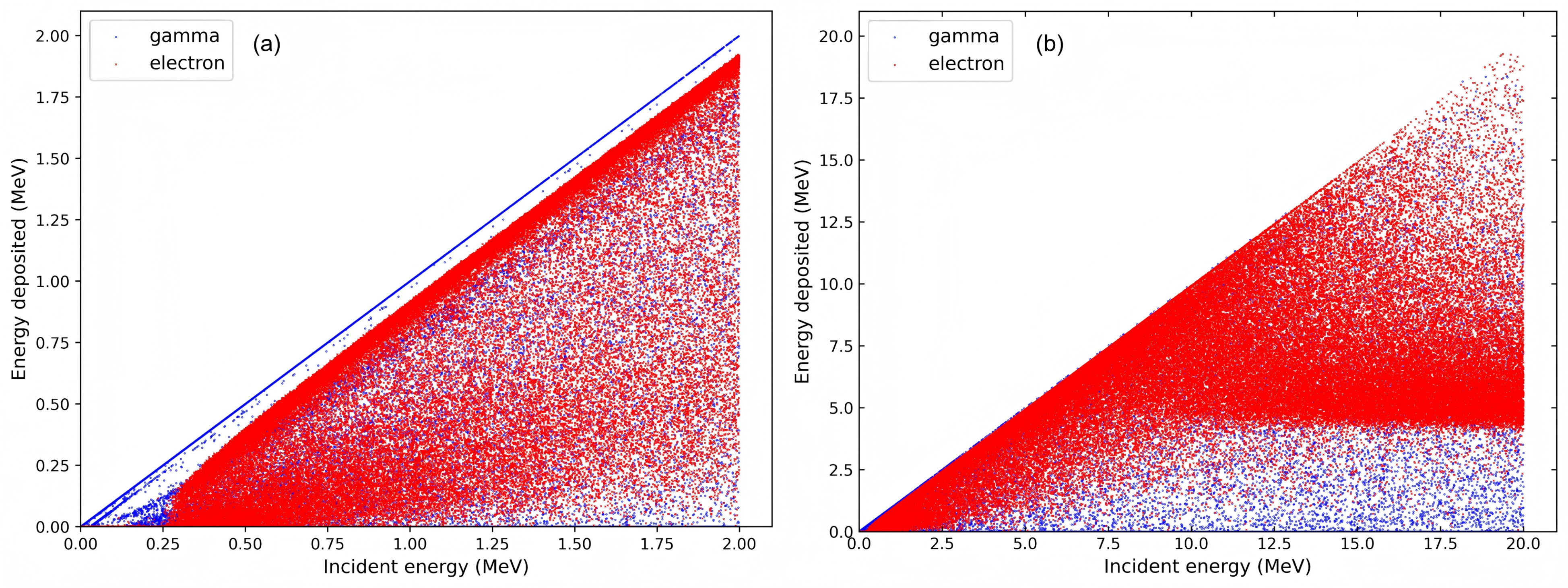}
\caption{Electron response matrix and gamma response matrix of the GTP for the narrow energy range of 0-2 MeV (a) and the wide energy range of 0-20 MeV (b).}
\label{fig:8}
\end{figure*}

Geant4, developed by the European Organization for Nuclear Research (CERN), is a Monte Carlo (MC) application package used for simulating particle transport through matter \cite{lu2022monte, huyan2018geant4}. In this study, we used version 11.0.3 of Geant4 to construct a GTP detector model identical to the one described in Section~\ref{chap:Design of Gamma-ray Transient Prob}. However, due to the blocking effect of the Be window and Teflon, the incident electrons were unable to deposit their full energy in the NaI(Tl) crystal. We first specifically investigated the deposited energy spectra of the NaI(Tl) crystal to incident electrons at 18 energy points within the 0.4-20 MeV range, both with and without the Be window and Teflon. 

In the Geant4 simulation, the GTP was placed in a vacuum environment, and 100,000 electrons were emitted from a distance of 50 mm from the center of the GTP at a 90° incident angle. We recorded the deposited energy in the NaI(Tl) crystal for each event. The partial results are shown in Fig.~\ref{fig:15}. The red curves represent the deposited energy spectra in the NaI(Tl) crystal without the Be window and Teflon, showing a distinct full-energy peak and a plateau caused by electron scattering. In contrast, the black curves represent the deposited energy spectra with the Be window and Teflon, showing a low-energy electron scattering platform and a peak at the most probable energy. As clearly shown in Fig.~\ref{fig:15}, the Be window and Teflon significantly absorb the electrons' energy. Only after nearly 100 keV of energy is absorbed can the electrons penetrate the Be window and Teflon to deposit energy in the NaI(Tl) crystal.

The energy transferred to the NaI(Tl) crystal after passing through the Be window and Teflon reflective material from a monoenergetic incident electron is not singular but forms a distribution. The interaction mechanisms between electrons and the NaI(Tl) crystal are diverse, mainly including three types: (1) multiple scattering and ionization, resulting in the deposition of the full energy; (2) partial energy deposition followed by scattering out of the NaI(Tl) crystal; (3) bremsstrahlung producing gamma photons, with three possible outcomes for these photons: (a) directly escaping the NaI(Tl) crystal, (b) undergoing a photoelectric cascade process that produces multiple Auger electrons and characteristic X-rays, or (c) undergoing Compton scattering, generating one or more Compton electrons, with secondary photons then undergoing a photoelectric cascade process.

From the simulation results, the deposited energy spectrum in NaI(Tl) crystal lacks a full-energy peak but features a continuous plateau caused by electron scattering and a most probable energy peak. The deposited energy spectra of NaI(Tl) crystal for different incident electron energies are plotted on the same graph for comparative analysis, as shown in Fig.~\ref{fig:11}. As the incident electron energy increases, the most probable energy peak detected by NaI(Tl) becomes more pronounced. However, at higher electron energies, the intensity and probability of bremsstrahlung increase, causing high-energy electrons to deposit only part of their energy, with the remainder escaping as gamma photons produced by bremsstrahlung. This explains why, as shown in Fig.~\ref{fig:11} (b), the count of events in the high-energy region of the NaI(Tl) deposition spectrum is relatively low for incident electron energy of 20 MeV.

We fitted the most probable energy peaks at 11 energy points within the 0.4-1.4 MeV range using the inverse-Landau function, as shown in Equation~\ref{eq:2}. In Equation~\ref{eq:2}, $MPV$ represents the most probable energy deposited by electrons in the NaI(Tl) crystal, $\eta$ denotes the width of the inverse-Landau distribution, and $A$ is the amplitude parameter. The fitting results for incident electron energies of 0.6 MeV and 1.2 MeV are shown in Figs.~\ref{fig:17} (a) and (b), respectively. Figure~\ref{fig:18} (a) illustrates the relationship between the most probable energy and the incident electron energy, which is fitted using a quadratic polynomial (Equation~\ref{eq:3}). In the subsequent Section~\ref{chap:Energy Response}, the most probable energy obtained from the Geant4 simulation will be used to ensure the accuracy of the energy response when presenting the energy-channel and energy-resolution relationships. The detection efficiency of the GTP is defined as the ratio of the number of events with non-zero deposited energy in the NaI(Tl) crystal to the number of incident electron events. Figure~\ref{fig:18} (b) shows the GTP detection efficiency in the energy range of 0.2-1.4 MeV. When the incident electron energy is below 250 keV, the detection efficiency is nearly zero, but when the incident energy exceeds 500 keV, the detection efficiency surpasses 90\%. For incident electron energies of 300 keV and 400 keV, approximately 24\% and 75\% of electron events, respectively, can penetrate the Be window and Teflon material to reach the NaI(Tl) crystal.

\begin{equation}\label{eq:2}
f(x;MPV,\eta,A)=A \cdot e^{-\frac{1}{2}(\frac{MPV-x}{\eta}+ e^{\frac{x-MPV}{\eta}})}
\end{equation}

\begin{equation}\label{eq:3}
f(x)=a_0+a_1 \cdot x+a_2 \cdot x^2
\end{equation}


To study the response over a continuous energy range, we also simulated the energy response matrix of the GTP to both electrons and gamma photons. The G4UniformRand() function in Geant4 was used to generate random energies within the ranges of 0-2 MeV and 0-20 MeV. At a distance of 50 mm from the GTP center, 100,000 electron events and 100,000 gamma events with these random energies were emitted at a 90° incident angle. Figure~\ref{fig:8} (a) shows the response matrix for electrons (red markers) and gamma rays (blue markers) with incident energies in the narrow range of 0-2 MeV. Significant detection counts for electrons were only observed when the incident electron energy exceeded 0.25 MeV. Due to the obstruction of the Be window and Teflon material, the maximum energy deposited by electrons differs consistently from the full-energy peak of gamma rays with the same incident energy. Figure~\ref{fig:8} (b) shows the response matrix for the wide energy range of 0-20 MeV. When the incident energy exceeds 15 MeV, scattering and bremsstrahlung reduce the number of counts in the high-energy region of the NaI(Tl) deposition spectrum. For gamma rays with excessively high energies, the probability of directly penetrating the GTP increases significantly, resulting in a sharp decrease in full-energy deposition events.

\section{Electron-Beam Testing Results}

This section presents the results of the ground-based electron performance tests for the GTP, focusing on two main aspects: the dead-time testing for electron beams in the 0.4-20 MeV range and the energy response to electron beams in the 0.4-1.4 MeV range.

\subsection{Dead-Time Testing}

\begin{figure*}[!htb]
\includegraphics[width=\hsize]
{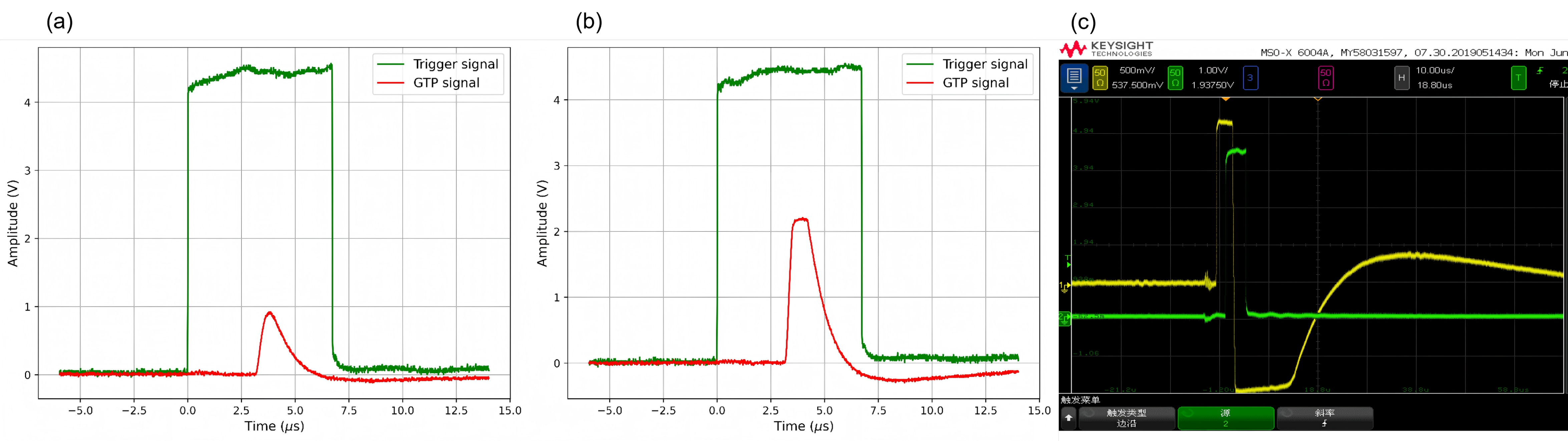}
\caption{Pulse waveform of the quasi-single electron beam on the GTP, which includes: normal event (a) and overflow events (b) and (c). Figure (c) shows a screenshot from the oscilloscope, where the yellow trace represents the overflow pulse signal generated by the 20 MeV electron beam on the GTP, while the green trace represents the trigger pulse signal.}
\label{fig:14}
\end{figure*}

In the experiments, we first tested the pulse shape and energy spectrum using the electron accelerator to investigate the GTP's electron response. Following the method described in Section~\ref{chap:High-energy Electron Accelerator Calibration Facility}, the GTP was installed in the vacuum chamber at the accelerator terminal, and the signal was connected to the data acquisition system outside the chamber via cables (Fig.~\ref{fig:2}). The chamber door was then closed, and a vacuum was created (Fig.~\ref{fig:3} (c)). The electron accelerator provided the beam and a trigger pulse, allowing the observation of the electron beam pulse signal from the GTP on an oscilloscope. Figure~\ref{fig:14} displays the pulse waveform of the quasi-single electron beam. Figure~\ref{fig:14} (a) shows the normal signal detected by the GTP, with a pulse width of 2-3 $\mu$s. To ensure the signal returns to the baseline, the dead time of the GTP for normal events is designed to be a fixed 4 $\mu$s. In Fig.~\ref{fig:14} (b) and (c), the pulse width exceeds 4 $\mu$s. These cases are referred to as overflow signals, in which only the dead time duration is recorded, while the amplitude is not. The designed dead time for such signals in the GTP is 70 $\mu$s.

\begin{figure*}[!htb]
\includegraphics[width=\hsize]
{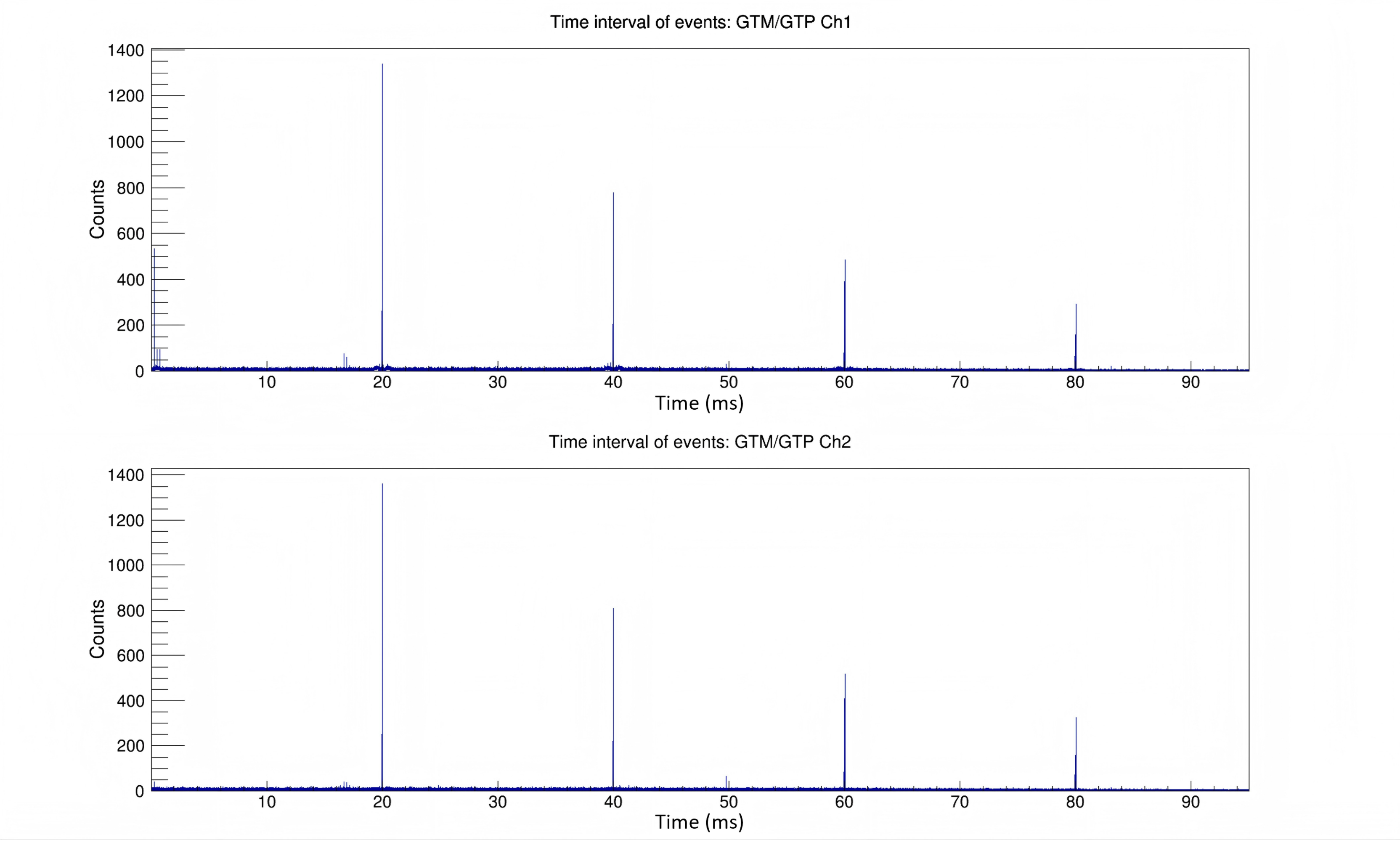}
\caption{Time interval spectrum of the selected 0.7 MeV electron beam at 50 Hz. Ch1 and Ch2 represent the two readout channels of the GTP.}
\label{fig:12}
\end{figure*}

For electron beams with energies exceeding 1.4 MeV, the signals detected by the GTP exceed the energy channel range. Therefore, we only test its dead time rather than the energy spectrum. When the GTP detects electrons greater than 1.4 MeV, it almost always produces overflow signals. The undershoot and baseline recovery time of overflow signals are primarily determined by the slew rate of the preamplifier chip. Figure~\ref{fig:14} (c) shows the overflow pulse signal generated by the 20 MeV electron beam on the GTP, with a dead time of approximately 70 $\mu$s, consistent with the designed value of the data acquisition system. The time interval spectrum of the GTP was analyzed using a 50 Hz, 0.7 MeV electron beam (Fig.~\ref{fig:12}). It was found that the time intervals of events detected by the GTP were multiples of 20 ms, indicating that the GTP’s time-recording capability is working correctly. The time interval of the accelerator electron beam is 20 ms, with a time jitter of less than 60 $\mu$s, while the GTP data acquisition system records with a time precision of approximately 100 ns.

\subsection{Energy Response}\label{chap:Energy Response}

\begin{figure*}[!htb]
\includegraphics[width=\hsize]
{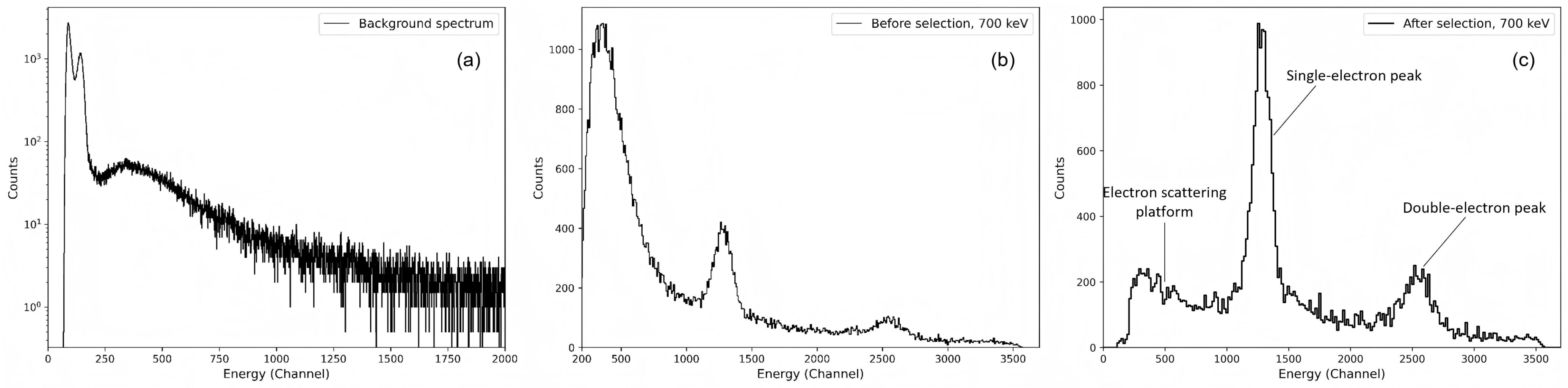}
\caption{(a) Background spectrum detected by the GTP. (b) Deposited energy spectrum before time interval selection of electron beam events. (c) Deposited energy spectrum after time interval selection of electron beam events.}
\label{fig:13} 
\end{figure*}

\begin{figure*}[!htb]
\includegraphics[width=\hsize]
{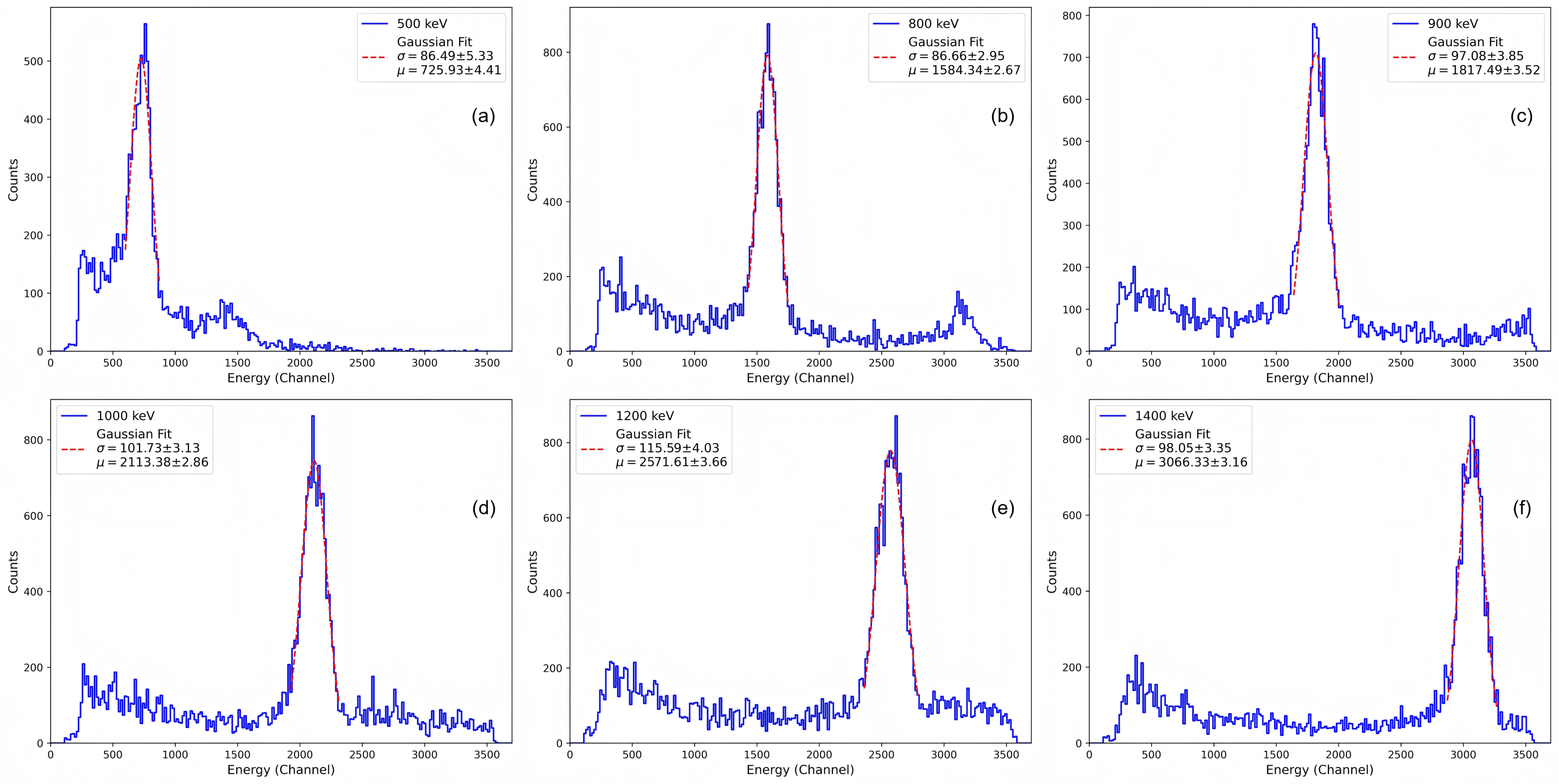}
\caption{Net deposited energy spectra of the GTP for 0.5 MeV (a), 0.8 MeV (b), 0.9 MeV (c), 1 MeV (d), 1.2 MeV (e), and 1.4 MeV (f) electron beams.}
\label{fig:9} 
\end{figure*}

\begin{figure*}[!htb]
\includegraphics[width=\hsize]
{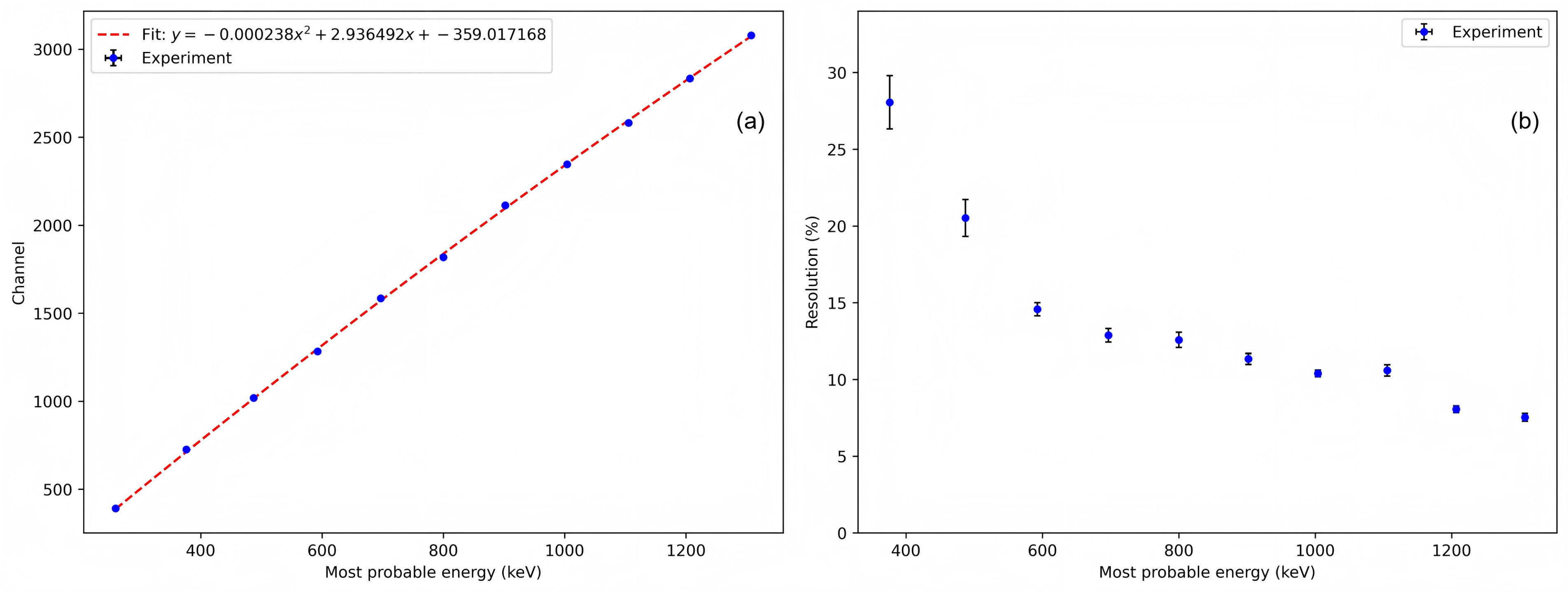}
\caption{(a) Energy-channel relationship of the GTP was established by fitting these data points with the quadratic polynomial. (b) Energy resolution of the GTP, with the uncertainties calculated using the error propagation formula.}
\label{fig:10}
\end{figure*}

We used the electron accelerator to test the energy response of the GTP for electron beams in the 0.4-1.4 MeV energy range. The experiment was conducted at 100 keV intervals, with a total of 11 tested energy points and at least 10,000 counts per point. During the experiment, at least five sets of background data were collected. Figure~\ref{fig:13} (a) shows a background spectrum detected by the GTP, where the 59.5 keV full-energy peak of the embedded radioactive source $^{241}$Am is clearly visible. The background data reveal that the noise level of the GTP is slightly elevated; however, it does not impact the electron tests. Additionally, the temperature of the GTP remains relatively stable.

However, due to the high gamma-ray background count rate in the environment and the relatively low count rate of the electron beam, the energy peak of the electrons was not distinct. To address this issue, we not only increased the testing duration (3600 s for each energy point) but also filtered electron events based on the time interval spectrum. Since the time interval between electron pulses is 20 ms, with a time jitter of less than 60 $\mu$s, we selected electron events with time intervals of 20 ms ± 0.06 ms, 40 ms ± 0.06 ms, 60 ms ± 0.06 ms, 80 ms ± 0.06 ms, and 100 ms ± 0.06 ms. Note that Fig.~\ref{fig:12} shows the time interval spectrum of a 0.7 MeV electron beam at 50 Hz after filtering. The interference from gamma events and noise has been clearly eliminated, and the electron beam events of interest for this study have been successfully selected. Figures~\ref{fig:13} (b) and (c) show the energy spectra of the GTP detector for a 0.7 MeV electron beam, before and after event filtering, respectively. After event filtering, the electron energy spectrum clearly reveals a dominant single-electron component with a secondary double-electron component, as well as a platform caused by electron scattering. The experimentally measured energy spectrum is consistent with the simulation results, both exhibiting distinct electron scattering platforms and energy deposition peaks.

For the data collected by the GTP in the 0.4-1.4 MeV energy range, the event selection described above was applied, and the same process was performed for the background data. By subtracting the background from the deposited electron energy spectrum, a filtered energy spectrum was obtained, referred to as the net deposited energy spectrum. Figure~\ref{fig:9} shows the net deposited energy spectra at partial energy points and the corresponding Gaussian fitting results (Equation~\ref{eq:4}). These fitting results, such as the centroids of energy peak ($\mu$) and standard deviations ($\sigma$), provide a basis for establishing the energy-channel (E-C) relationship and energy resolution of the GTP for electrons. Figure~\ref{fig:10} (a) presents the E-C relationship of the GTP for electrons, along with the curve fitted using a quadratic polynomial (Equation~\ref{eq:3}). Figure~\ref{fig:10} (b) shows the energy resolution of the GTP. Here, the energy resolution is represented by the full width at half maximum (FWHM = 2.355 · $\sigma$), and its uncertainty is calculated using the error propagation formula.

\begin{equation}\label{eq:4}
f(x) = A \cdot e^{-\frac{(x-\mu)^2}{2\sigma^2}}
\end{equation}

It is important to note that electrons deposit energy while passing through the Be window and Teflon reflective materials, resulting in a deposited energy in the NaI(Tl) crystal that is lower than the beam energy. Therefore, the horizontal axis in Fig.~\ref{fig:10} needs to be corrected to the most probable energy of electrons, as determined by the Geant4 simulation described in Section~\ref{chap:Electron Energy Response Simulation}. In Fig.~\ref{fig:10} (a), the lowest incident energy is 400 keV, corresponding to a most probable energy of 259.8 keV in the GTP, while the highest incident energy is 1400 keV, corresponding to a most probable energy of 1307.4 keV. Figure~\ref{fig:10} (a) shows that the GTP's response to 400-1400 keV electrons is normal, with excellent linearity, as expected in the design. The electron energy spectrum measured at an incident energy of 400 keV is incomplete. Figure~\ref{fig:10} (b) shows only the energy resolution for incident energies in the range of 500-1400 keV. Due to the detection threshold and the limitations of the data acquisition system's energy channels, the GTP can detect energies in the range of 118-3566 channels. Based on the fitted E-C relationship curve (Fig.~\ref{fig:10} (a)), the corresponding range of most probable energies is calculated to be 165-1525 keV. Finally, using the relationship between most probable energy and incident energy obtained from Geant4 simulations (Fig.~\ref{fig:18} (a)), the GTP's detectable energy range for electrons is calculated to be 310-1629 keV.







\section{Summary}

The GTM payload is a novel all-sky monitor specifically designed for gamma-ray detection in the 20 keV to 1 MeV range, which was launched into the Distant Retrograde Orbit (DRO) in 2024. During its in-orbit operation, the GTM encounters strong electron and proton activity in the magnetotail region on a monthly basis. In this study, we conducted detailed performance testing of the GTP using the IHEP Electron-Beam Facility. The IHEP Electron-Beam Facility is a high-energy electron accelerator with continuously adjustable energy, ideally suited for testing the GTP's electron energy response. We primarily tested the GTP's dead time for electron signals in the range of 0.4-20 MeV and the deposited energy spectrum in the range of 0.4-1.4 MeV. The test results show that the GTP’s dead time for normal signals is less than 4 $\mu$s, while for overflow signals it is approximately 70 $\mu$s, consistent with the design specifications. The GTP recorded overflow events accurately and its performance met expectations. We also performed Geant4 simulations to study the GTP’s electron energy spectrum with and without obstructive materials. The simulations revealed that electrons can only deposit energy in the GTP if they carry energy greater than 250 keV to penetrate the Be window and Teflon materials. The simulation results were consistent with the experimental data. Considering the E-C relationship, detection thresholds, and channel limitations of the data acquisition system, we concluded that the GTP’s detection energy range for electrons is from 310 keV to 1629 keV. The ground-based electron tests enhances the mass model of the detector and verifies its response in the high-energy range, laying the foundation for establishing a calibration database and representing a critical step in the development and operation of the GTM payload. Testing the efficiency of the GTP using electron beams proved challenging due to the electron scattering issue inherent in NaI(Tl) detectors. For the GTP’s electron detection efficiency, we provided simulation results. Additionally, future work will involve proton testing using a backup GTP. Currently, the data acquisition program cannot provide a direct relationship between signal width and energy, and we recommend adding this functionality in future engineering projects to extend the energy measurement range.


\section{Acknowledgments}

This work is supported by the Strategic Priority Research Program of the Chinese Academy of Sciences (Grant No. XDA30050100) and the National Natural Science Foundation of China (Grant Nos. 12173038, 11775251, 12273042, and 12075258). The GECAM (Huairou-1) mission was funded by the Strategic Priority Research Program on Space Science (XDA15360000) of the Chinese Academy of Sciences (CAS). We thank the staff of the Shandong Institute of Aerospace Electronic Technology, National Institute of Metrology, for significantly helping in the development and ground calibration tests of the GTM. We also thank the DRO team.




\section*{CONFLICT OF INTEREST}

The authors declare that they have no competing interests.

\bibliography{mybibfile}

\end{document}